 \documentclass[final,5p,times,twocolumn]{elsarticle}

\usepackage{amssymb}

\journal{Physica C: Superconductivity and its applications}

\begin{document}

\begin{frontmatter}



\title{An effect of the uniaxial strain on the temperature of Bose-Einstein condensation of the intersite bipolarons}


\author[*,**]{B. Ya. Yavidov}
\ead{yavidov@inp.uz}
\address[*]{Phase Transitions Physics Laboratory, Institute of Nuclear Physics, 100214 Ulughbek, Tashkent, Uzbekistan}
\address[**]{Department of Polymer Physics, National University of Uzbekistan named after Mirzo Ulughbek, 100174 Tashkent, Uzbekistan}
\begin{abstract}
We have studied an effect of uniaxial strain to the temperature of
Bose-Einstein condensation of intersite bipolarons within the
framework of Extended Holstein-Hubbard model. Uniaxial lattice
strains are taken into an account by introducing a generalized
density-displacement type force for electron-lattice interaction.
Associating the superconducting critical temperature $T_c$ with the
temperature of Bose-Einstein condensation $T_{BEC}$ of intersite
bipolarons we have calculated strain derivatives of $T_{BEC}$ and
satisfactorily explained the results of the experiments on La-based
high-$T_c$ films.
\end{abstract}


\end{frontmatter}


\section{Introduction}
The first observation of high-$T_c$ superconductivity in cuprates
(copper oxide La-Ba-Cu-O compound) was reported almost quarter
century ago \cite{bed-mull}. Since then an intensive both
theoretical and experimental research studies have been carried out
on revealing the microscopic origin of the high-$T_c$
superconductivity. In despite of decades of investigations at
present there is no consent about the microscopic mechanism that
bounds electrons (holes) into pairs, and consequently leads to
high-$T_c$ superconductivity. Normal state of cuprates also exhibit
a variety of phase states. The early and subsequent studies show
that an electronic subsystem in the cuprates is strongly correlated
and role of the electron-lattice interactions are significant
\cite{bar-bishop,gunnar,mish}. It has been understood that a number
of physical parameters of the cuprates as well as the critical
temperature ($T_c$) strongly depend on a variety of factors as
oxygen content, doping level, applied pressure or strain for films,
and etc. Similar to some low-$T_c$ superconductors an applied
external pressure increases the value of $T_c$ of copper based
oxides. The highest critical temperature $T_c\simeq 164$ K was
obtained for HgBa$_2$Ca$_2$Cu$_3$O$_{8+x}$ under hydrostatic
pressure \cite{gao} ($T_c\simeq 133$ K at ambient pressure
\cite{schilling}). So far an external pressure remains as powerful
tool that affects all physical parameters of a sample simultaneously
and influences the value of critical temperature. However, the
influence of an external pressure to the cuprates is not a
straightforward. In particular, it is the case when considering the
dependence of a critical temperature on applied external pressure.
The same is true for the films grown on various substrates. In the
latter case pressure originates from lattice mismatch between
substrate and film. The lattice mismatch causes a microscopic strain
which in its turn, on the analogy of pressure, vary the critical
temperature. Due to quasi two dimensional (2D) lattice structure and
strong anisotropy of the properties of cuprates the dependence of
the critical temperature on uniaxial strain (pressure) is not
simple. For many cuprate compounds compressive uniaxial strain
(pressure) in the CuO$_2$ plane (along ${\bf a}$- or ${\bf b}$-
axes) enhances $T_c$, while compressive uniaxial strain (pressure)
along ${\bf c}$- axis reduces $T_c$
\cite{nohara,gugenberger-nakamura,meingast-walker,fukamachi,chen_x,sato-naito,
locquet}. There are contradictorily reports in the cases of
YBa$_2$Cu$_3$O$_{7-\delta}$ and GdBa$_2$Cu$_3$O$_{7-\delta}$ where
compressive strain (pressure) in the CuO$_2$ plane gives rise to
increase \cite{belenky-greene,meingast} or decrease
\cite{budko-schuller,welp} of the critical temperature. Moreover,
the sign of the pressure derivatives of $T_c$ along ${\bf a}$ and
${\bf b}$ axes are different for YBa$_2$Cu$_3$O$_{7-\delta}$
\cite{kraut-wuhl}. Uniaxial pressure study of electron-doped cuprate
Nd$_{1.84}$Ce$_{0.16}$CuO$_4$ has shown that dependence of the
critical temperature of the electron doped compound are the same
with those of hole doped counterparts \cite{kaga-takagi}.

There are a few theoretical studies of the effect of an uniaxial
strain (pressure) on the critical temperature
\cite{goddard,q.p.li,klein-simanovsky,chen-habermeier,ovidko}.
Goddard estimated the effect of the pressure on the critical
temperature of the optimally doped La$_2$CuO$_4$ within the magnon
pairing  model \cite{goddard}. The analytical formula for $T_c$ was
obtained in the weak coupling BCS approach. The coupling parameters
of the model enter to the formula of $T_c$ and depend on the orbital
overlaps. An applied pressure decreases the distance between the
centers of orbits, and this in its turn leads to the exponential
increase of coupling parameters. In this way $T_c$ is affected by
the pressure (strain). The uniaxial pressure results of Goddard
along ${\bf a}$- and ${\bf b}$- axes are in good agreement with the
results of Ref. \cite{gugenberger-nakamura}. Meanwhile, the
prediction of the model with regard to the effect of the pressure
along ${\bf c}$ axis is smaller than those results of Ref.
\cite{gugenberger-nakamura} by order. The effect of the uniaxial
stress to the critical temperature $T_c$ of YBa$_2$Cu$_3$O$_7$
cuprate was studied by Li in Ref.\cite{q.p.li} in terms of the
anisotropy of the density of states at the Fermi energy. Using a
simple picture of van Hove singularity of high-$T_c$ superconductors
Li was able satisfactorily reproduce the results of Ref. \cite{welp}
and qualitatively explain the anisotropy of strain derivatives of
the critical temperature. The main features of the work
\cite{q.p.li} are: (i) consideration is pure two dimensional; (ii)
weak coupling BCS formula for $T_c$ or spin fluctuation induced
superconductivity formula of Pines is applied to study the
dependence of $T_c$ on uniaxial stress. In reality the cuprates have
quasi two dimensional structure and it is believed that BCS approach
hardly applicable to the cuprates. In addition, Ref. \cite{q.p.li}
did't discuss an effect of stress along ${\bf c}$ axis to $T_c$.
Klein and Simanovsky studied in Ref. \cite{klein-simanovsky}
anisotropic pressure dependence of $T_c$ of YBa$_2$Cu$_3$O$_7$
within the framework of tunneling unit model.  It was assumed that
the conduction electrons scatter in so called tunneling units
(double-well potentials) that are oriented along ${\bf a}$ axis.The
uniaxial strain along ${\bf a}$ axis was associated with the change
of microscopic width of the tunneling unit (the distance between the
two wells). Within the model it is appears that the change of $T_c$
in direct proportion to the relative change of microscopic width of
tunneling unit. The pressure derivatives of $T_c$ along both ${\bf
a}$ and ${\bf b}$ axes was calculated in close agreement with the
experimental results. The strain derivative of $T_c$ along ${\bf b}$
axis was interrelated with those of along ${\bf a}$ axis. According
Ref.\cite{klein-simanovsky} uniaxial compressive pressure in the
${\bf b}$ direction decreases the distances in the ${\bf b}$
direction, while the those increases the distances in the ${\bf a}$
direction. However, the effect of the uniaxial pressure (strain)
along ${\bf a}$ or ${\bf b}$ axis to the expansion of the distances
along ${\bf c}$ axis and contribution arising from this
interrelation to $T_c$  was not discussed. Chen et al studied in
Ref. \cite{chen-habermeier} an anisotropy of $T_c$ under uniaxial
pressure for optimally doped YBa$_2$Cu$_3$O$_{7-\delta}$ within the
framework of anisotropic $t-J$ model. Though the theoretical results
of Ref. \cite{chen-habermeier} are in good agreement with
experiments, the reduction of $T_c$ under uniaxial pressure along
${\bf c}$-axis was mainly attributed due to the pressure-induced
increase of holes density in CuO$_2$- plane (so-called pressure
induced charge transfer model (PICTM) \cite{millis-rabe}). In the
meanwhile, the same PICTM fails to explain the uniaxial pressure
effects in the $ab$- plane of YBa$_2$Cu$_3$O$_{7-\delta}$
\cite{kraut-wuhl}.

The cumulative effect of the strains in both ${\bf a}$ and ${\bf b}$ axes to $T_c$ of Bi$_2$Sr$_2$CaCu$_2$O$_x$, the strains in both ${\bf a}$ and ${\bf c}$ (${\bf b}$ and ${\bf c}$) to $T_c$ of YBa$_2$Cu$_3$O$_{7-x}$ were discussed in Ref. \cite{ovidko}. Such type of consideration is necessarily  when one studying the dependence of $T_c$ of cuprate films grown in various substrates. Indeed, due to mismatch of the lattice parameters (periods) of film and substrate the grown film is exposed to stress along all axes simultaneously. If lattice period of a substrate is smaller (lager) than those of a film then the grown film experiences compressive (tensile) stress in the plane parallel to the boundary plane of film and substrate. The compressive (tensile) stress of the film along $ab$, $bc$ and $ac$ planes gives rise to lengthening (shortening) of the distances along ${\bf c}$, ${\bf a}$ and ${\bf b}$ axes, respectively. The importance of account of the stress along ${\bf c}$ axis was emphasized in Ref. \cite{sato-naito}. Moreover, the highest $T_c=49$ K was achieved for La$_{1.9}$Sr$_{0.1}$CuO$_4$ film mainly due to cumulative effect of all axes stresses \cite{locquet}.

The experimental data obtained so far and tested models for
different cuprate compounds suggest that the phenomenon is complex
and differ for various compounds. Nevertheless there are common
features too. As a consequence at present one experiences a need of
a model that able to explain all experimental data from universal
points of view or to point out, at least, the most relevant
contributions. From theoretical point of view the works
\cite{goddard,q.p.li,klein-simanovsky,ovidko} study the effect of
uniaxial strain on $T_c$ totally ignoring quasi two dimensionality
of the structure of cuprates. In addition, they ignore the most
relevant electron-phonon interaction. The crucial role of
electron-phonon interaction in cuprates comes from different
experiments \cite{bar-bishop}. It is well established that
electron-phonon interaction in cuprates is strong and (bi)polaron is
formed in the cuprates \cite{asa-mott-report,salje-asa}. These
issues are not discussed in the Refs.
\cite{goddard,q.p.li,klein-simanovsky,chen-habermeier,ovidko}. In
present paper we will try to fill this blank. We study an effect of
uniaxial strain to $T_c$ within the framework of the extended
Holstein-Hubbard model (EHHM). EHHM is believed appropriate model
for study strongly correlated electron-phonon systems. An estimation
of $T_c$ within the model will be given for a chain model of
cuprates at strong coupling limit and nonadiabatic regime, and
discussion of the obtained results with regard to La-based
high-$T_c$ films will be done. Here we associate $T_c$ with the
Bose-Einstein condensation temperature of intersite bipolarons which
form due to strong electron-phonon interaction.

\begin{figure}[tbp]
\begin{center}
\includegraphics[angle=-0,width=0.5\textwidth]{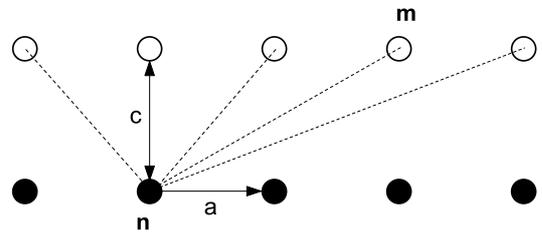} \vskip -0.5mm
\end{center}
\caption{Chain model of cuprates is represented by the extended Holstein model in which an electron performs hopping motion in a one-dimensional chain of the molecules (lower chain) and interacts with all ions of an upper chain via a density-displacement type force $f({\bf n}-{\bf m})$ (dashed lines). The distances between the chains and between the ions are assumed equal to 1.}
\end{figure}

\section{Hamiltonian for a strained lattice}
The systems that have strongly correlated electron and strongly
coupled electron-phonon subsystems can be described by means of
extended Holstein-Hubbard model which proved to be a more realistic
model for cuprates. Hamiltonian of the system is given by
\cite{akfcm}
\begin{equation}\label{1}
H=H_{e}+H_{ph}+H_{V}+H_{e-ph},
\end{equation}
where
\begin{equation}\label{2}
H_{e}=\sum_{{\bf n}\neq {\bf n'}}T({\bf n}-{\bf n'})c^{\dag}_{\bf n}c_{\bf n'},
\end{equation}
describes hopping of electrons between adjacent sites,
\begin{equation}\label{3}
H_{ph}=\sum_{{\bf q},{\alpha}}\hbar\omega_{{\bf
q}\alpha}\left(d^{\dag}_{\bf q\alpha}d_{{\bf q}\alpha}+1/2\right),
\end{equation}
is Hamiltonain of phonon system,
\begin{equation}\label{4}
H_{V}=\sum_{{\bf n}\neq {\bf n'}}V_{C}({\bf n}-{\bf n'})c^{\dag}_{\bf n}c_{\bf n}c^{\dag}_{\bf n'}c_{\bf n'},
\end{equation}
is Hamiltonian of interacting particles on sites $\bf n$ and $\bf n'$ via Coulomb forces,
\begin{equation}\label{5}
H_{e-ph}=\sum_{{\bf n}{\bf m}\alpha}f_{{\bf m}\alpha}({\bf
n})c^{\dag}_{\bf n}c_{\bf n}\xi_{{\bf m}\alpha}
\end{equation}
is Hamiltonian of electron-phonon interaction. Here $T(\bf n-\bf
n')$ is transfer integral of electron form site $\bf n$ to site $\bf
n'$, $c^{\dag}_{\bf n}(c_{\bf n})$ is creation (annihilation)
operator of electron on site $\bf n$, $d^{\dagger}_{\bf
q\alpha}(d_{\bf q\alpha})$ is creation (annihilation) operator of
phonon with $\alpha$ ($\alpha=x,y,z$) polarization and wave vector
$\bf q$, $\omega_{{\bf q}\alpha}$ is the phonon's frequency,
$V_{C}(\bf n-\bf n')$ is Coulomb potential energy of two electrons
located at sites $\bf n$ and $\bf n'$, $f_{\bf m\alpha}(\bf n)$ is
"density-displacement" type coupling force of an electron on site
${\bf n}$ with the ions on site ${\bf m}$, $\xi_{{\bf m}\alpha}$ is
normal coordinate of ion vibrations on site ${\bf m}$ which
expresses through phonon creation and destruction operators as
\begin{equation}\label{6}
    \xi_{{\bf m}\alpha}=\sum_{\bf q}\left(\sqrt{\frac{\hbar}{2NM\omega_{{\bf
    q}\alpha}}}e^{i{\bf qm}}d^{\dagger}_{{\bf
    q}\alpha}+h.c.\right).
\end{equation}
Here $N$ is number of sites and $M$ is ion mass.

EHHM enables one to take into account electron correlations and long
range feature of electron-phonon interaction. Many physical
properties of cuprates were satisfactorily explained within the EHHM
(see for example \cite{asa-b2003}) as well as the possibility of
formation of intersite bipolarons with $s$-, $p$- and $d$- type wave
functions \cite{physc-yav}. The possibility of formation of magnetic
bipolaron with $d$- type wave function is also shown in Ref.
\cite{vid} within the framework of $t-J$ model.

Though model Hamiltonian Eq.(1) enables one to study polaron and
bipolaron formation in a discrete lattice, here for the sake of
simplicity estimation of polaron's mass will be done without the
term Eq.(4) (i.e. within extended Holstein model) and for the
dispersionless phonons. In strong electron-phonon coupling limit and
nonadiabatic regime use of the standard procedures such as
Lang-Firsov transformation \cite{lang-fir} eliminates
electron-phonon interaction term (5). Subsequent perturbation
expansion of the transformed Hamiltonian $H_e$ with respect to
parameter $\lambda=E_p/2T(a)$ ($E_p$ is polaronic shift) and
estimation of polaron's renormalized mass for a lattice in Fig.1
yields $m_p/m^{\ast}=\exp[g^2]$ \cite{alekor} (see also
\cite{asaya}), where
\begin{equation}\label{6}
    g^2=\frac{1}{2M\hbar\omega^3}\sum_{\bf m}[f^2_{{\bf m}\alpha}({\bf n})-f_{{\bf m}\alpha}({\bf n})f_{{\bf m}\alpha}({\bf n}+{\bf a})].
\end{equation}
and $m^{\ast}=\hbar^2/2T(a)a^2$ is the bare band mass. Lattice of Figure 1 was introduced by Alexandrov and Kornilovitch in Ref. \cite{alekor}
 in order to mimic an interaction of a hole on CuO$_2$ plane with the vibrations of {\it apical} ions in the cuprates. Convincing evidence for a
 such coupling of in-plane holes with the $c$- axis polarized vibrations of apical ions comes from many experiments (see for example \cite{timusk}).
  Therefore from here we will omit index $\alpha$ and consider only that component of the electron-lattice force which represents an interaction of
  a hole on CuO$_2$ plane with the $c$- axis polarized apical oxygen vibrations. In addition, in order to consider stress of a lattice and its
  influence to the (bi)polaron mass, and consequently to the temperature of Bose-Einstein condensation of intersite bipolarons an analytical expression
\begin{equation}\label{7}
  f_{\bf m}({\bf n})=\frac{\kappa c(1-\varepsilon_c)}{[|({\bf n}-{\bf m})(1-\varepsilon_{a})|^2+(c(1-\varepsilon_c))^2]^{3/2}}
\end{equation}
will be accepted for the density-displacement type force. Here $\kappa$ is some coefficient, $\varepsilon_a$ and $\varepsilon_c$ are lattice strains
along ${\bf a}$ and ${\bf c}$ axes, respectively. $|{\bf n}-{\bf m}|$ is measured in units of lattice constant $|{\bf a}|=1$. The lattice strains
defined as $\varepsilon_a=(a_{unst}-a_{str})/a_{unst}$ and $\varepsilon_c=(c_{unst}-c_{str})/c_{unst}$, where subscripts {\it unstr} and {\it str}
stand for unstrained and strained, respectively. Eq.(7) is generalization of the force considered in Ref. \cite{alekor} (see Eq.(9) there)
 and allows one to interrelate the temperature of Bose-Einstein condensation of the intersite bipolarons with the lattice strains through the
 mass of intersite bipolaron. Indeed, it has been shown that within the model Eq.(1) intersite bipolaron tunnel in the first order of polaron
 tunneling and its mass has the same order as polaron mass \cite{akfcm}. For the sake of simplicity we suppose that intersite bipolarons form an
 ideal gas of charged carriers and mass of bipolaron is $m_{bp}=2m_p$ (this point does not lead to loose of generality). Then the temperature of
 Bose-Einstein condensation of the intersite bipolarons defines as
\begin{equation}\label{8}
 T_{BEC}=\frac{3.31\hbar^2n^{2/3}}{2k_Bm^{\ast}}e^{-g^2}.
\end{equation}
Here $k_B$ is Boltzmann constant and $n$ is density of intersite bipolarons.
The Eqs.(6), (7) and (8) are the main analytical results of the paper, according which discussion of experimental data for La-based high-$T_c$ thin films will be done in the next section.

\section{Results and discussion}
The Eq.(8) expresses $T_{BEC}$ through two basic parameters of a
system: (i) the density of intersite bipolarons $n$ and (ii) the
exponent $g^2$ of the polaron mass enhancement. Here we discuss the
possibility of application of the Eq.(8) to thin films of La-based
high-$T_c$ cuprates grown on different substrates. This choice is
quite simple and reasonably. In contrast with other cuprates
La$_{2-x}$Sr$_x$CuO$_4$ has somewhat a simplest crystal structure
and serves as a test material for a variety of theoretical models.
In addition, it is commonly believed that pressure induced charge
transfer model (PICTM) \cite{millis-rabe} does not work in
La$_{2-x}$Sr$_x$CuO$_4$ compound due to lack of chain structures. In
these circumstances Eq.(8) allows one to study the dependence of
$T_{BEC}$ on lattice strains $\varepsilon_a$ or $\varepsilon_c$ at
constant $n$. This dependence is, of course, originated obviously
from polaronic effects. We have calculated the values of $T_{BEC}$
as a function of the strains along ${\bf a}$ axis $\varepsilon_a$
and ${\bf c}$ axis $\varepsilon_c$ for the model lattice given in
Fig.1. Here we put $n=1\cdot 10^{21}$ sm$^{-3}$ and
$\kappa^2/(2M\hbar\omega^3)=8.51$ in order to coincide $T_{BEC}$ at
the absence of the strain with the bulk value of $T_c=25$ K of
La$_{1.9}$Sr$_{0.1}$CuO$_4$ \cite{locquet}. The results are
visualized in Fig.2. As one can see from Fig.2(a) compressive strain
along ${\bf a}$ axis gives rise to increase of $T_{BEC}$, while
those along ${\bf c}$ axis acts on the contrary, Fig.2(b). Uniaxial
strain derivatives of $T_{BEC}$ of the model lattice (Fig.1) are:
$\partial T_{BEC}/\partial\varepsilon_a\approx 112$ K and $\partial
T_{BEC}/\partial\varepsilon_c\approx -450$ K.

\begin{figure}[tbp]
\begin{center}
\includegraphics[angle=-0,width=0.5\textwidth]{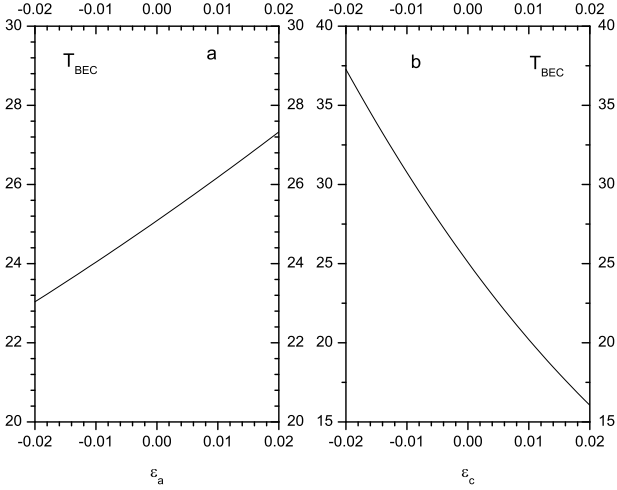} \vskip -0.5mm
\end{center}
\caption{The temperature of Bose-Einstein condensation of the
intersite bipolarons as a function of strains along ${\bf a}$ axis
$\varepsilon_a$ (a) and along ${\bf c}$ axis $\varepsilon_c$ (b).
Here we put $\kappa^2/(2M\hbar\omega^3)=8.51$ in order to coincide
$T_{BEC}$ at $\varepsilon_i=0$ with the bulk value of $T_c=25$ K of
La$_{1.9}$Sr$_{0.1}$CuO$_4$ \cite{locquet}.}
\end{figure}

\begin{figure}[tbp]
\begin{center}
\includegraphics[angle=-0,width=0.5\textwidth]{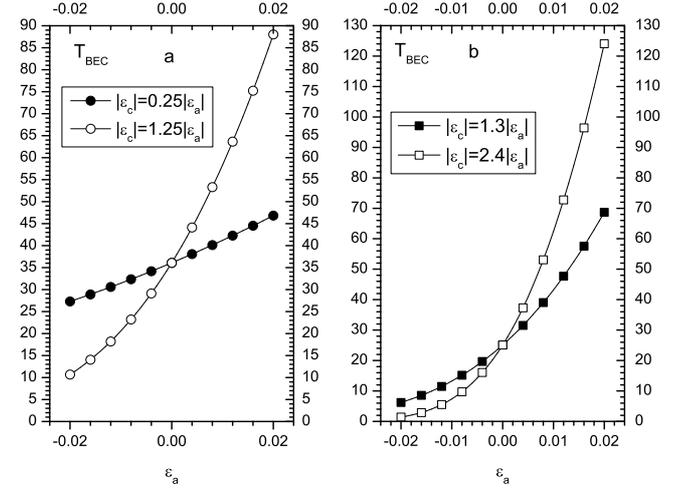} \vskip -0.5mm
\end{center}
\caption{The temperature of Bose-Einstein condensation of the
intersite bipolarons as a function of strain along ${\bf a}$ axis
$\varepsilon_a$. In Fig.3(a) and Fig.3(b) $T_{BEC}$ are plotted for
two films reported in Refs. \cite{sato-naito} and \cite{locquet},
respectively. In accordance with these we put
$\kappa^2/(2M\hbar\omega^3)=7.77$ and
$\kappa^2/(2M\hbar\omega^3)=8.51$ in Fig.3(a) and Fig.3(b),
respectively, in order to coincide $T_{BEC}$ at $\varepsilon_i=0$
with the value of $T_c$ of unstrained sample of the experiment.}
\end{figure}

The strain derivatives of $T_c$ $\partial T_c/\partial\varepsilon_i$
($i=a,b,c$) were reported for a series of high-quality single
crystals of La$_{2-x}$Sr$_x$CuO$_4$ from high-resolution dilatometry
experiments in Ref. \cite{gugenberger-nakamura}. According to this
report $\partial T_c/\partial\varepsilon_i$ depends on doping level
and lies in range of $70\div 470$ K, $280\div 580$ K and $-2440\div
-1090$ K  for $i=a$, $i=b$ and $i=c$, respectively. Comparison of
these data show that the value of $\partial
T_{BEC}/\partial\varepsilon_a\approx 112$ K of our model lattice
lies in the range of $\partial T_c/\partial\varepsilon_a$. However,
$\partial T_{BEC}/\partial\varepsilon_a$ lies well below of the
range of $\partial T_c/\partial\varepsilon_b$ and absolute value
$\partial T_{BEC}/\partial\varepsilon_c$ is approximately four times
smaller than mean absolute value of $\partial
T_c/\partial\varepsilon_c$. Another experiment reports about
uniaxial strain derivatives of $T_c$ of
La$_{1.91}$Sr$_{0.09}$CuO$_4$ \cite{nohara} in which $\partial
T_c/\partial\varepsilon_{ab}=284$ K and $\partial
T_c/\partial\varepsilon_c=-851$ K. One can easily extract the value
of $\partial T_c/\partial\varepsilon_{i}$ ($i=a,b$) from $\partial
T_c/\partial\varepsilon_{ab}$  if one assumes that CuO$_2$ composed
of isotropic square lattice, and the strains along ${\bf a}$ and
${\bf b}$ axes are approximately the same
$\varepsilon_a\approx\varepsilon_b$ and
$\varepsilon_{ab}=\varepsilon_a+\varepsilon_b$. Then one finds
$\partial T_c/\partial\varepsilon_i=142$ K ($i=a,b$) which is close
to our $\partial T_{BEC}/\partial\varepsilon_a$. Meanwhile our
$\partial T_{BEC}/\partial\varepsilon_c$ is still lesser than
$\partial T_c/\partial\varepsilon_c$ of Ref. \cite{nohara}, now by
two times. The reasons for a such discrepancies will be discussed
below. Though strain derivatives of $T_{BEC}$ are differ from those
of bulk samples we will see below that they can explain the strain
effects to $T_c$ of La-based high-$T_c$ films.

Let us now turn our attention to the results of two particular
experiments of Refs. \cite{sato-naito,locquet}. The experiments
report about the growth of thin La-based high-$T_c$ films on various
substrates and an influence of lattice mismatch between film and
substrate to the $T_c$.

As reported in \cite{sato-naito} Sato and Naito have grown of (001)
oriented La$_{1.85}$Sr$_{0.15}$CuO$_4$ (LSCO) thin films on
LaSrAlO$_4$ (LSAO) and SrTiO$_3$ (STO) substrates by reactive
co-evaporation method. Unstrained in-plane lattice parameters of
these LSCO, LSAO and STO are 3.777 {\AA}, 3.756 {\AA} and 3.905
{\AA}, respectively. They were able to obtain LSCO film on LSAO
substrate with $T_c$ as high as 44 K, which is approximately 7 K
higher than that for single LSCO crystal under ambient pressure. For
the films on STO substrates they obtained $T_c=29$ K. X-ray
diffraction measurements of LSCO thin films grown on LSAO (STO)
substrate clearly showed that the lattice parameter of CuO$_2$-
plane is compressed by 0.4$\%$ (expanded by 1.6$\%$), while ${\bf
c}$-axis is expanded by 0.5$\%$ (compressed by 0.4$\%$). As a matter
of recorded results from structural analysis Sato and Naito
concluded that increase (decrease) of $T_c$  is directly related to
both the compression (tension) of CuO$_2$- plane by the strain
generated by lattice mismatch and the expansion (shortening) of the
${\bf c}$- axis due to the Poisson effect. The ratio of the strain
along ${\bf c}$- axis to the strain of CuO$_2$-plane lattice
parameter were: $|\varepsilon_c|/|\varepsilon_a|=1.25$ and
$|\varepsilon_c|/|\varepsilon_a|=0.25$ for the films grown on LSAO
and STO substrates, respectively. We have performed calculation of
$T_{BEC}$ according to the Eqs.(6)-(8) taking into account the above
relations $|\varepsilon_c|/|\varepsilon_a|$. The results are
presented graphically in Fig.3(a). For the compressive strain
$\varepsilon_a=0.4\%$ our model yields $T_{BEC}=44.1$ K which is
very close to the observed value of $T_c=44$ K for LSCO film grown
on LSAO substrate. In the meantime for the tensile strain
$\varepsilon_a=-1.6\%$ our model yields $T_{BEC}=28.9$ K which is
also very close to the observed value of $T_c=29$ K for LSCO film
grown on STO substrate. One should notice that there is an excellent
agreement between our results and the experimental results of Ref.
\cite{sato-naito}.

Locquet et al reported growth of La$_{1.9}$Sr$_{0.1}$CuO$_4$ (LSCO)
thin films on LaSrAlO$_4$ (LSAO) and SrTiO$_3$ (STO) substrates by
the block-by-block molecular epitaxial method in Ref.
\cite{locquet}. The $\theta-2\theta$ X-ray measurements revealed
that the grown films strongly strained both in CuO$_2$-plane and out
of the plane. The film grown on LSAO substrate was found strained
with parameters $\varepsilon_{ab}=0.63\%$ and
$\varepsilon_c=-0.76\%$. The same parameters for the film grown on
STO substrate were found as $\varepsilon_{ab}=-0.54\%$ and
$\varepsilon_c=0.35\%$. The ratio of the strains
$|\varepsilon_c|/|\varepsilon_a|$ for the films grown on LSAO and
STO substrates are 1.2 and 0.65, respectively. Takin into account
that $\varepsilon_{ab}\approx 2\varepsilon_a$ one rewrites the same
relations as $|\varepsilon_c|/|\varepsilon_a|=2.4$ and
$|\varepsilon_c|/|\varepsilon_a|=1.3$ for the films grown on LSAO
and STO substrates, respectively. The results of the calculation of
$T_{BEC}$ for our lattice with the same relations are given in
Fig.3(b). For the compressive lattice strain $\varepsilon_a=0.63\%$
and the relation $|\varepsilon_c|/|\varepsilon_a|=2.4$ our model
yields the value of $T_{BEC}= 46$ K, while for the tensile strain
$\varepsilon_a=-0.54\%$ and the relation
$|\varepsilon_c|/|\varepsilon_a|=1.3$ one obtains $T_{BEC}\approx
18$ K. As one can see our result $T_{BEC}= 46$ K is in reasonable
agreement with the value of $T_c=49$ K for the film on LSAO
substrate. However, there is some discrepancy of the value of
$T_{BEC}\approx 18$ K from observed value of $T_c=10$ K for the film
on STO substrate.

Let's now briefly discuss some aspects of our model and its results.
As pointed out above our model takes into account the most relevant
interaction in cuprates which is electron-phonon interaction. We
were able satisfactorily explain the $T_c$ of La-based high-$T_c$
films within the framework of EHHM and in the assumption of
formation of ideal Bose-gas of intersite bipolarons. Some
discrepancies of our results from the experimental results might be
as result of several factors: (i)  the simplest of our model lattice
under study. In reality one should consider more complex structures;
(ii) the choice of the analytical formula for the
density-displacement type electron-lattice force; (iii) an
assumption that intersite bipolarons form an ideal Bose-gas. In
reality due to other factors there are may be deviation from an
ideality thus forming nonideal Bose-gas or Bose-liquid; (iv)
superconductivity of the La-based films may be due to not the only
electron-phonon interaction, but has contributions from other
interactions as well. These factors suggest performance of more
comprehensive research on the studied problem. Even though we have
considered a very simple lattice model its common features are
consistent qualitatively with many experimental observations.
Indeed, in our model compressive strain in $ab$-plane of high-$T_c$
cuprates gives rise an increase of $T_c$, while that of along
$c$-axis acts contrary. Such a type variations of $T_c$ with respect
to the strains were observed in many La-based cuprates
\cite{suzuki,cieplak-lindenfeld,kao-mannaerts,tabata-kawai,nohara-suzuki,hanaguri-kojima,motoi-sakudo}.
This feature is common both to films
\cite{suzuki,cieplak-lindenfeld,kao-mannaerts,tabata-kawai} and bulk
samples \cite{nohara-suzuki,hanaguri-kojima,motoi-sakudo}. Though
our model allows one estimate $T_{BEC}$ at any $\varepsilon_i$ there
are might be technological limits in obtaining of films with a large
value of strains induced by lattice mismatch. It is also worthwhile
to notice that our consideration are limited to the films of optimal
thickness. Because of generality of our approach one might want to
extend the model to other cuprates. In general, there is no
restrictions for that. However, in each case specific features of a
compound (film or bulk sample) must be taken into an account.

\section{Conclusion}
In this paper we have considered an effect of uniaxial strain
(pressure or stress) to the superconducting critical temperature of
cuprates. A brief survey of both experimental and theoretical works
on the effect of the uniaxial pressure (strain or stress) to $T_c$
of cuprates (bulk and film samples) reveal that the phenomenon is
complex and there are contradictory reports too. However, it is
emphasized that there are common features relevant to all of them.
In particular, in the experiments with La-based high-$T_c$ cuprates
compressive pressure (strain) both along ${\bf a}$ and ${\bf b}$
axes gives rise to increase of $T_c$, while that along ${\bf c}$
axis gives rise to decrease of $T_c$. On the other hand, there are
convincing evidences of importance of electron correlations,
electron-lattice interactions and bipolaron formation in the
cuprates. Within the framework of extended Holstein-Hubbard model we
introduced a generalized density-displacement type force for the
electron-lattice interaction which takes into account lattice
strain. We were able express bipolaron mass through the lattice
strains. In the assumption that intersite bipolarons form an ideal
Bose-gas, and associating the superconducting critical temperature
$T_c$ with the temperature of Bose-Einstein condensation of
intersite bipolarons we directly relate $T_{BEC}$ with lattice
strain $\varepsilon_i$ ($i=a,b,c$). Operating on a simple model
lattice introduced early in Ref. \cite{alekor} we have calculated
strain derivatives of $T_{BEC}$. Further we have tried to explain
the effect of the uniaxial strain to the superconducting critical
temperature of La-based high-$T_c$ films. Here we limit ourselves to
two particular experiments of Refs. \cite{sato-naito,locquet}. Our
calculated results for $T_{BEC}$ are in close agreement with the
results of experiments of Refs. \cite{sato-naito,locquet} for $T_c$.
A brief discussion is made of the reasons of small discrepancies of
our results from experimental ones.

The author acknowledges financial support from the Foundation of the
Fundamental Research of Uzbek Academy of Sciences (grant no.
${\Phi}$A-${\Phi}$2-${\Phi}$070).

\end{document}